\documentclass{article}

\usepackage{amsfonts}
\usepackage{epsfig}

\begin{document}

\title{\textsc{Ghetto of Venice: Access to the Target Node and the Random Target Access Time}}

\vspace{1cm}

\author{ D. Volchenkov and Ph. Blanchard
\vspace{0.5cm}\\
{\it  BiBoS, University Bielefeld, Postfach 100131,}\\
{\it D-33501, Bielefeld, Germany} \\
{\it Phone: +49 (0)521 / 106-2972 } \\
{\it Fax: +49 (0)521 / 106-6455 } \\
{\it E-Mail: VOLCHENK@Physik.Uni-Bielefeld.DE}}
\large

\date{\today}
\maketitle

\begin{abstract}
Random walks defined on undirected graphs
assign the absolute scores to all nodes  based on
the quality of  path they provide for random walkers.
In city space syntax, the notion of
segregation acquires a statistical interpretation
with respect to random walks.
We analyze the spatial network
 of Venetian canals and detect its most segregated part
which can be identified with canals
adjacent to the Ghetto of Venice.
\end{abstract}

\vspace{0.5cm}

\leftline{\textbf{ PACS codes: 89.75.Fb, 89.75.-k, 89.90.+n} }
 \vspace{0.5cm}

\leftline{\textbf{ Keywords: Complex networks; traffic equilibrium; city space syntax.} }

\section{Introduction}
\label{sec:Introduction}
 \noindent

The spatial network
 of Venice that stretches across 122 small
 islands between which the canals serve the function of roads is constituted by
96 canals.
In March, 1516 the Government of the Serenissima Repubblica issued
special laws, and the first {\it Ghetto} of Europe was instituted.
It was an area where Jews were forced to live and which they could
not leave from sunset to dawn. The Ghetto existed for more than two
and a half centuries, until Napoleon conquered Venice and finally
opened and eliminated every gate (1797).

 The phenomenon of clustering of minorities, especially that of newly
 arrived immigrants, is well documented \cite{Wirth} (the reference appears
 in \cite{Vaughan01}).
Clustering is considering to be beneficial for mutual support and
for the sustenance of cultural and religious activities. At the
same time, clustering and the subsequent physical segregation of
minority groups would cause their economic marginalization.
The spatial analysis of the immigrant
quarters \cite{Vaughan01}  and the
study of London's changes over 100 years \cite{Vaughan02}
shows that they were significantly more
segregated from the neighboring areas, in particular, the number
of street turning away from the quarters to the city centers were
found to be less than in the other inner-city areas being
usually socially barricaded by the railways, canals and industries.

It has been suggested \cite{Language} that space structure and its
impact on  movement are critical to the link between the built
environment and its social functioning. Spatial structures creating a local situation
in which there is no relation between movements inside the spatial
pattern and outside it and the lack of natural space occupancy
become associated with the social misuse of the structurally
abandoned spaces.
It is well known that the urban layout effects on the spatial
distribution of crime
\cite{Crime}. It has been noted that crime seems
to be highest where the urban grid is most broken
 up (in effect creating most local segregation), and lowest
 where the lines are longest, and in fact most integrated.
  The linear routes through the estate have least burglary,
  and the most broken up, locally enclosed spaces the most
  \cite{Language}.

In the present paper, we analyze the integration/segregation
statistics of the Venetian canal network by means of random walks.
In the forthcoming section (Sec.~\ref{sec:Space_Syntax}), we
discuss the role of city space syntax for  attaining  traffic
equilibrium in the city transport network. Then, in
Sec.~\ref{sec:Study_of_undirected_graphs_by_means_of_random_walks},
we show that random walks establish the Euclidean space structure
on undirected graph, in which distances and angles have the clear
statistical interpretations well known in quantitative theory of
random walks. In Secs.~\ref{sec:Ghetto} and
\ref{sec:AccessTarget}, we analyze Venetian space syntax and
detect the Ghetto of Venice as the most segregated part of the
city canal network. Then, we conclude in the last section.

\section{City space syntax}
\label{sec:Space_Syntax}
\noindent

Studies of urban networks have a long history.
In most of researches
devoted to the improving  of transport routes, the optimization
of power grids, and the pedestrian movement surveys
 the relationships between certain components of the urban texture
  are often measured along streets and routes considered as edges
  of a planar graph, while the traffic end points and street
  junctions are treated as nodes. Such a  {\it primary} graph
  representation of urban networks is grounded on relations
  between junctions through the segments of streets. The usual
   city map based on Euclidean geometry can be considered as an
   example of primary city graphs.

The notion of traffic equilibrium had been introduced by J.G. Wardrop in
 \cite{Wardrop:1952} and then generalized in \cite{Beckmann:1956} to a
 fundamental concept of   network equilibrium. Given a connected undirected
  graph $G(V,E)$, in which $V$ is the
set of nodes and $E$ is the set of edges, we can define the traffic
 volume $f: E\to(0,\infty[$ through every edge $e\in E$. It then follows
  from the Perron-Frobenius theorem that the linear equation
\begin{equation}
\label{Lim_equilibrium}
f(e)\,=\, \sum_{e'\,\in\, E}\,f(e')\,\exp\left(\,-h\,\ell\left(e'\right)\,\right)
\end{equation}
has a unique positive solution $f(e)>0$, for every edge $e\in E$, for a
fixed positive constant $h>0$ and a chosen set of positive  {\it metric
 length} distances $\ell(e)>0$. This solution is naturally identified
 with the traffic equilibrium state of the transport network defined on
  $G$, in which the permeability of edges depends upon their lengths.
The parameter $h$  is called the volume entropy of the graph $G$, while
 the volume of $G$ is defined as the sum $\mathrm{Vol}(G)=\frac 12\sum_{e\in E}\ell(e)$.

The degree of a node $v\in V$ is the number of its neighbors in
$G$, $\deg(v)=k_v$. It has been shown in \cite{Lim:2005} that
among all undirected connected graphs of normalized volume,
$\mathrm{Vol}(G)=1$, which are not cycles and $k_v\ne 1$ for all
nodes,
 the minimal possible value of the volume entropy,
\begin{equation}
\label{min_h}
\min(h)=\frac 12\sum_{v\in V}k_v\,\log\left(k_v-1\right)
\end{equation}
 is attained
for the length distances
\begin{equation}
\label{ell_min}
\ell(e)\,=\,\frac {\log\left(\left(k_{i(e)}-1\right)
\left(k_{t(e)}-1\right)\right)}{2\,\min(h)},
\end{equation}
where $i(e)\in V$ and $t(e)\in V$ are the initial and terminal vertices
of the edge $e\in E$ respectively. It is then obvious that substituting
(\ref{min_h}) and (\ref{ell_min}) into (\ref{Lim_equilibrium}) the
operator $\exp\left(-h \ell(e)\right)$ is given by a
 symmetric Markov transition operator,
\begin{equation}
\label{Markov_transition}
f(e)\,=\, \sum_{e'\,\in\, E}\,\frac{f(e')}{\sqrt{\left(k_{i(e)}-1\right)
\left(k_{t(e)}-1\right)}},
\end{equation}
which rather describes time reversible random walks over edges than over
nodes. The flows satisfying (\ref{Lim_equilibrium}) with the operator
(\ref{Markov_transition}) meet the mass conservation property,
$\sum_{i\sim j}f_{ij}=\pi_j$, $\sum_{j\in V}\pi_j=1$.
The Eq.(\ref{Markov_transition}) unveils the indispensable role Markov's
 chains defined on edges play in equilibrium traffic modelling and exposes
 the degrees of nodes as a key determinant of the transport networks properties.

 Wardrop's traffic
 equilibrium \cite{Wardrop:1952} is strongly tied to the human apprehension of space since
  it is required that all travellers have enough knowledge of the
  transport network they use. The human perception of places is not
  an entirely Euclidean one, but are rather related to the perceiving
  of the vista spaces (streets and squares) as single units and of the
   understanding of the topological relationships between these vista spaces,
   \cite{Kuipers}. Decomposition of city space into a complete set of
   intersecting vista spaces produces a spatial network which we call
   the {\it dual} graph representation of a city. Therein, the relations
    between streets treated as nodes are traced through their  junctions
     considered as edges.

Dual city graphs are extensively investigated within the concept of
{\it space syntax}, a theory developed in the late 1970s, that seeks to
reveal the mutual effects of complex spatial urban networks on society
and vice versa, \cite{Hillier:1984,Hillier:1999}.  Spatial perception
 that shapes peoples understanding of how a place is organized  determines
eventually the pattern of local movement which is predicted
by the space syntax method with surprising accuracy \cite{Penn:2001}.
The robustness of agreement between centrality of spaces
and rush hour movement rates is now supported by a
number of similar studies of pedestrian movement in different
parts of the world and in an everyday commercial work of the {\it Space Syntax Ltd.},
 \cite{Read1997}.

\section{Study of undirected graphs by means of random walks}
\label{sec:Study_of_undirected_graphs_by_means_of_random_walks}
\noindent

 The issues of global connectivity of
finite graphs and
accessibility of their nodes
have always been the classical fields
of researches in graph theory.
Any graph representation  naturally arises as an outcome of categorization,
when we abstract a real world system by eliminating all but one of its features
 and by the grouping of things (or places) sharing a common attribute by
 classes or categories. For instance, the common attribute of all open spaces
 in city space syntax is that we can move through them. All elements called
 nodes that fall into one and the same group $V$ are considered as essentially
 identical; permutations of them within the  group are of no consequence.
The symmetric group $\mathbb{S}_{N}$ consisting of all permutations of $N$
elements
($N$ being the cardinality of the set $V$) constitute the symmetry group of $V$.
If we denote by $E\subseteq V\times V$ the set of ordered pairs of nodes called
edges, then  a graph is a map $G(V,E): E \to K\subseteq\mathbb{R}_{\,+}$
(we suppose that the graph has no multiple edges).

The nodes of $G(V,E)$ are weighted with respect to some  {\it  measure}
 $m=\sum_{i\in V} m_i \delta_i,$ specified by a set of positive numbers $m_i> 0$.
  The space $\ell^2(m)$ of square-assumable functions with respect to the
   measure $m$ is the {\it Hilbert space} $\mathcal{H}$ (a complete inner
   product  space). Among all linear operators defined on $\mathcal{H}$,
   those  {\it invariant} under the permutations of nodes are of essential
   interest since they reflect the symmetry of the graph. Although there are
   infinitely many such operators, only those which maintain  conservation
   of a quantity may describe a physical process. The Markov transition
   operators which share the property of  {\it probability conservation}
   considered in theory of random walks on graphs are among them.
Laplace operators describing diffusions on graphs meet the {\it mean value}
 property ({\it mass conservation}); they give another example \cite{Smola2003}
  studied in spectral graph theory.

Being defined on connected undirected graphs, a Markov transition operator $T$ has a unique
{\it equilibrium} state $\pi$ (a stationary distribution of random walks)
 such that $\pi T=\pi$ and $\pi=\lim_{t\to\infty}\,\sigma\, T^{\,t}$ for
  any density $\sigma\in \mathcal{H}$ ($\sigma_i\geq 0$,
$\sum_{i\in V} \sigma_i=1$). There is a unique measure
$m_\pi =\sum_{i\,\in\, V} \pi_i\delta_i $ related to the stationary distribution $\pi$
with respect to which the Markov operator $T$
is {\it self-adjoint},
\begin{equation}
\label{s_a_analogue}
\widehat{T}=\,\frac 12
\left( \pi^{1/2}\,\, T\,\,
\pi^{-1/2}+\pi^{-1/2}\,\, T^\top\,\,
 \pi^{1/2}\right),
\end{equation}
where $T^\top$ is the adjoint operator. The orthonormal ordered set of real
 eigenvectors $\psi_i$, $i=1\ldots N$, of the symmetric operator $\widehat{T}$
  defines a basis in  $\mathcal{H}$. In quantitative theory of
  random walks  defined on graphs \cite{Lovasz:1993,Aldous} and in spectral
  graph theory \cite{Chung:1997}, the properties of graphs are studied in
   connection with the  eigenvalues and eigenvectors of self-adjoint operators
   defined on them. In particular, the symmetric transition operator defined on
   undirected graphs is $\widehat{T_{ij}}=1/\sqrt{k_ik_j}$. Its first eigenvector
    $\psi_1$ belonging to the largest eigenvalue $\mu_1=1$,
\begin{equation}
\label{psi_1}
\psi_1
\,\widehat{ T}\, =\,
\psi_1,
\quad \psi_{1,i}^2\,=\,\pi_i,
\end{equation}
describes the {\it local} property of nodes (connectivity), $\pi_i=k_i/2M,$
 where $2M=\sum_{i\in V} k_i$, while the remaining eigenvectors
 $\left\{\,\psi_s\,\right\}_{s=2}^N$ belonging to the eigenvalues
  $1>\mu_2\geq\ldots\mu_N\geq -1$ describe the {\it global} connectedness of the graph.

Markov's symmetric transition operator $\widehat{T}$  defines a {\it projection}
 of any density $\sigma\in \mathcal{H}$ on the eigenvector $\psi_1$ of the
  stationary distribution $\pi$,
\begin{equation}
\label{project}
\sigma\,\widehat{T}\,
=\,\psi_1 + \sigma^\bot\,\widehat{T},\quad \sigma^\bot\,=\,\sigma-\psi_1,
\end{equation}
in which $\sigma^{\bot}$ is the vector belonging to the orthogonal complement of 
$\psi_1$.
Thus, it is clear that any two densities $\sigma,\rho\,\in\,\mathcal{H}$ differ
 with respect to random walks only by their dynamical components, $(\sigma-\rho)\,
 \widehat{T}^t\,=\,(\sigma^\bot -\rho^\bot)\,
\widehat{T}^t$ for all $t\,>\,0$.
Therefore, we can define a
distance between any two densities which they acquire  with respect to random walks by
\begin{equation}
\label{distance}
\left\|\,\sigma-\rho\,\right\|^2_T\, =
\, \sum_{t\,\geq\, 0}\, \left\langle\, \sigma-\rho\,\left|T^t
\right|\, \sigma-\rho\,\right\rangle.
\end{equation}
 or, in the spectral form,
\begin{equation}
\label{spectral_dist}
\left\|\sigma-\rho\right\|^2_T =
\sum_{t\,\geq 0} \sum_{s=2}^N \mu^t_s \left\langle
\sigma-\rho|\psi_s\right\rangle\!\left\langle \psi_s
| \sigma-\rho\,\right\rangle
  =  \sum_{s=2}^N\frac{\left\langle \sigma-\rho|
 \psi_s\right\rangle\!\left\langle \psi_s
| \sigma-\rho\right\rangle}{1-\mu_s},
\end{equation}
where we have used  Dirac's  bra-ket notations especially
convenient in working with inner products and
rank-one
operators in Hilbert space.

If we introduce in $\mathcal{H}(V)$ a new inner product by
\begin{equation}
\label{inner-product}
\left(\,\sigma,\rho\,\right)_{T}
\,= \, \sum_{t\,\geq\, 0}\, \sum_{s=2}^N
\,\frac{\,\left\langle\, \sigma\,|\,\psi_s\,\right\rangle\!
\left\langle\,\psi_s\,|\,\rho \right\rangle}{\,1\,-\,\mu_s\,}
\end{equation}
for all  $\sigma,\rho\,\in\, \mathcal{H}(V),$
then (\ref{spectral_dist}) is nothing else as
\begin{equation}
\label{spectr-dist2}
\left\|\,\sigma-\rho\,\right\|^2_T\, =
\left\|\,\sigma\,\right\|^2_T +
\left\|\,\rho\,\right\|^2_T  -
2 \left(\,\sigma,\rho\,\right)_T,
\end{equation}
 where
\begin{equation}
\label{sqaured_norm}
\left\|\, \sigma\,\right\|^2_T\,=\,
\,\sum_{s=2}^N \,\frac{\left\langle\, \sigma\,|\,\psi_s\,\right\rangle\!
\left\langle\,\psi_s\,|\,\sigma\, \right\rangle}{\,1\,-\,\mu_s\,}
\end{equation}
being the squared norm of  $\sigma\,\in\, \mathcal{H}(V)$ with respect to
random walks.
We finish the description of the $(N-1)$-dimensional Euclidean
space structure associated to random walks by mentioning that
given two densities $\sigma,\rho\,\in\, \mathcal{H}(V),$ the
angle between them can be introduced in the standard way,
\begin{equation}
\label{angle}
\cos \,\angle \left(\rho,\sigma\right)=
\frac{\,\left(\,\sigma,\rho\,\right)_T\,}
{\left\|\,\sigma\,\right\|_T\,\left\|\,\rho\,\right\|_T}.
\end{equation}
Random walks embed connected undirected graphs into Euclidean
space. This embedding can be used in order to compare
 nodes
and to retrace
 the optimal coarse-graining
representations.

Namely,  let us consider
the density $\delta_i$ which equals 1 at
the node $i\,\in\, V$ and zero for all other nodes. It takes form
$\upsilon_i\,=\,\pi^{-1/2}_i\,\delta_i$ with respect to the measure
$m_\pi$. Then, the squared norm of  $\upsilon_i$ is given by
\begin{equation}
\label{norm_node}
\left\|\,\upsilon_i\,\right\|_T^2\, =\,{\frac 1{\pi_i}\,\sum_{s=2}^N\,
\frac{\,\psi^2_{s,i}\,}{\,1-\mu_s\,}},
\end{equation}
where $\psi_{s,i}$ is the $i^{\mathrm{th}}$-component of the
eigenvector $\psi_s$. In  theory of random walks \cite{Lovasz:1993},
the quantity (\ref{norm_node}) expresses the {\it access time} to a target node
quantifying the expected number
of  steps
required for a random walker
to reach the node
$i\in V$ starting from an
arbitrary
node  chosen randomly
among all other
nodes  with respect to
the stationary distribution $\pi$.

The Euclidean distance between any two nodes of the graph $G$
established by random walks,
\begin{equation}
\label{commute}
\begin{array}{lcl}
K_{i,j}&=&\left\|\, \upsilon_i-\upsilon_j\,\right\|^2_T\\
 &=& H_{ij}+H_{ji},
\end{array}
\end{equation}
is known as the {\it commute time} in   theory of
random walks and  equals to the expected number of steps required
for a random walker starting at $i\,\in\, V$ to visit $j\,\in\, V$
and then to return to $i$ again,  \cite{Lovasz:1993}. The expected
number of steps a random walker  reaches $j$ if starts from $i$ is
called the {\it first hitting time} (or the {\it access time})
\cite{Lovasz:1993},
\begin{equation}
\label{access_time}
\begin{array}{lcl}
H_{ij} & = & \left\|\, \upsilon_j\,\right\|^2_T\,-\,\left(\upsilon_i,\upsilon_j\right)_T \\
 &= & \,\sum_{k=2}^N\,\frac 1{\,1-\mu_k\,}\,
\left(\frac{\,\psi^2_{k,j}\,}{\,\pi_j\,}-
\frac{\,\psi_{k,i}\,\psi_{k,j}\,}{\sqrt{\,\pi_i\,\pi_j\,}}\right).
\end{array}
\end{equation}
The cosine of an angle calculated in accordance to
 (\ref{angle}) has the structure of
Pearson's coefficient of linear correlations
 that reveals it's natural
statistical interpretation.
Correlation properties of flows
of random walkers
passing by different paths
 have been remained beyond the scope of
previous  studies devoted to complex
networks and random walks on graphs.
The notion of angle between any two nodes in the
graph arises naturally as soon as we
become interested in
the strength and direction of
a linear relationship between
two random variables,
the flows of random walks moving through them.
If the cosine of an angle (\ref{angle}) is 1
(zero angles),
there is an increasing linear relationship
between the flows of random walks through both nodes.
Otherwise, if it is close to -1 ($\pi$ angle),
  there is
a decreasing linear relationship.
The  correlation is 0 ($\pi/2$ angle)
if the variables are linearly independent.
It is important to mention that
 as usual the correlation between nodes
does not necessary imply a direct causal
relationship (an immediate connection)
between them.

In the forthcoming sections, we study the average of the
first access time $H_{ij}$ (\ref{access_time}) with respect to
its first and second indices.

\section{Detection of ghettos by access to a target node}
\label{sec:Ghetto}
\noindent

The Euclidean structure introduced on undirected graphs by random walks
can be used in order to investigate city space syntax.
Various properties of access times
have been recently studied in
 concern with the traffic flow
forecasting \cite{Sun},  in order
  to model the wireless
terminal movements in a cellular wireless
  network \cite{Jabbari},
in a statistical test for the presence
   of a random walk component
in the repeat sales price index
   models in house prices \cite{Hill},
in the growth modelling
   of urban conglomerations \cite{Pica},
and in many other
   works where random walks
have been considered directly on
   the city maps and physical landscapes.
In contrast to all previous studies,
 we use random
 walks in order to investigate
 the  morphology of
  urban textures described by
 dual city graphs. In the context of traditional space syntax
studies, the value of access time (the Euclidean norm, (\ref{norm_node}))
calculated for a node of a dual graph
can be naturally interpreted as
the  expected number of random elementary navigation
actions (i.e., the random turns at the
junctions between axial lines)
required to reach the certain open
space in the city.

The access to a target node introduced in theory
of random walks \cite{Lovasz:1993}
is the important {\it global} characteristic of the
{\it node} which is equal to the average of $H_{ji}$ (\ref{access_time})
 over its {\it first} index.
It quantifies the expected number of  steps required for a random walker
to reach the node starting from an arbitrary
node  chosen randomly among all other nodes of the network with respect to
the stationary distribution $\pi$,
\begin{equation}
\label{sverage1}
\sum_{j\,\in\, V}\,
\pi_j\,H_{ji}\, = \, \frac{1}{\pi_i}\,\sum_{k=2}^N\,\frac{\,\psi^2_{k,i}\,}{\,1-\mu_k\,}\,
= \, \left\|\,\upsilon_i\,\right\|^2_T.
\end{equation}
Being a global characteristic of a node in the graph, the access
(\ref{sverage1}) assigns the absolute scores to all nodes in the
network based on the quality of  path they provide for random
walks. In urban spatial networks, the value given by
(\ref{sverage1}) varies strongly from one open space to another:
the norm of a space (street, square,
 or canal) that can be easily  reached (just in a few random syntactic steps)
from any other space in the city  is minimal, while it could be very large for a
statistically segregated street.
\begin{figure}[ht]
 \noindent
\begin{center}
\epsfig{file=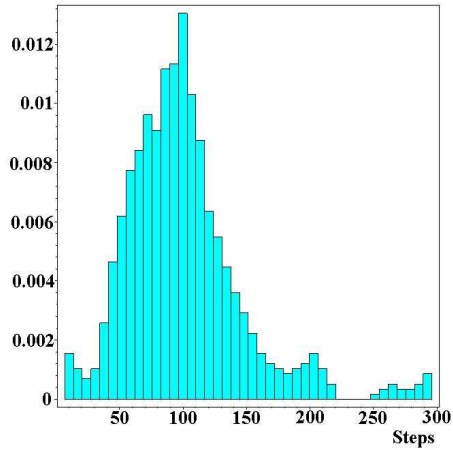,  angle= 0,width =6cm, height =6cm}
  \end{center}
\caption{\small Access spectrum of Venice canals. The  histogram  of
  norms acquired by the Venetian canals with respect to random walks. }
\label{Fig2_04}
\end{figure}
The relatively isolated groups of nodes and bottlenecks
 can be visually detected from
distribution histograms of the Euclidean norms
(see Fig.~\ref{Fig2_04}). In particular, it would
facilitate the detection of {\it ghettos} and urban
sprawl\footnote{Urban sprawl is the spreading out of a city and
its suburbs at the fringe of an urban area.}. In
Fig.~\ref{Fig2_04}, we have presented the  histogram of empirical
distribution of mean access times for the dual graph of Venetian
canals. It is interesting to note that the utmost peak on the
histogram correspondent from 250 to 300 random syntactic steps
represents the canals of the Cannaregio district
surrounding the quarter of Venetian Ghetto. From 16$^{\mathrm{th}}$ century,
 the quarter had been enlarged
  later to cover the neighboring Ghetto Vecchio and the
  Ghetto Nuovissimo. As a result a specific Ghetto canal subnetwork
  arose in Venice  weakly connected to the main canals.
\begin{figure}[ht]
 \noindent
\begin{center}
\epsfig{file=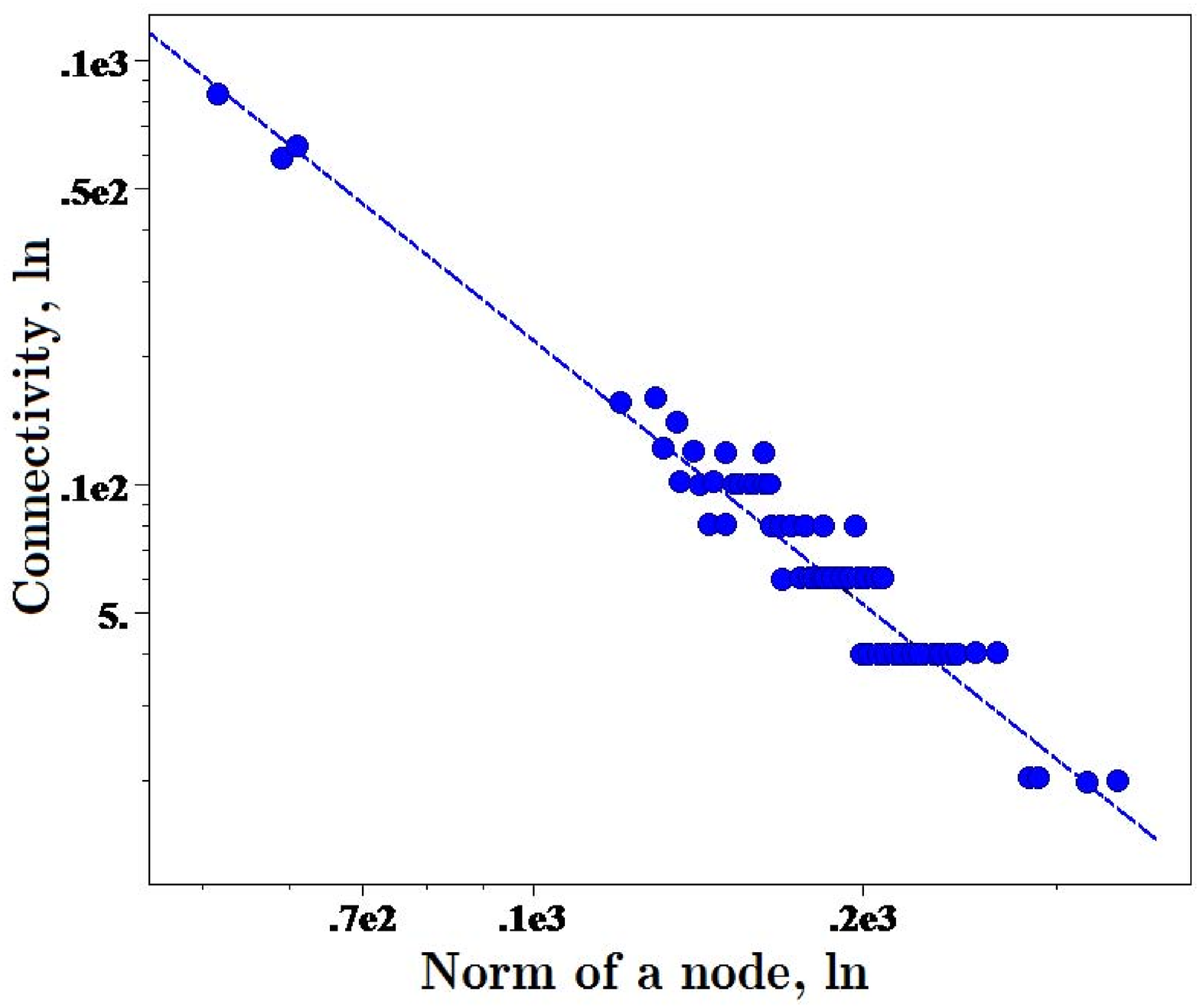,  angle= 0,width =6cm, height =5.5cm}
  \end{center}
\caption{\small The scatter plot of connectivity vs. the norm of a node in
the dual graph representation of 96 Venetian canals. The slope of the regression
 line equals 2.07.}
\label{Fig2_004a}
\end{figure}
The norm (\ref{norm_node}) which a node acquires in regard to random walks is a natural
statistical centrality measure of the vertex within a graph.
Probably, the most important message of space syntax theory is that
the local property (connectivity) of city spaces (streets and squares)
and their global configuration variable (centrality) are positively
 related in a city, and this part-whole relationship known as
 {\it intelligibility} \cite{Hillier:1999} is a key determinant
 of human behaviors in urban environments. An adequate
 level of intelligibility has been found to encourage
  peoples way-finding abilities. Intelligibility of Venetian
  canal network reveals itself quantitatively in the scaling of the
   norms of nodes with connectivity shown in Fig.~\ref{Fig2_004a}.
The three utmost points in the left upper part of the graph
displayed on Fig.~\ref{Fig2_004a} represent the most intelligible
structure in the Venetian canal network formed by the  Venetian
Lagoon, the Giudecca canal, and the Grand Canal. The four points
characterized by the worse accessibility levels delineate the
Ghetto canal subnetwork segregated statistically from the rest of
the network.

\section{Random target access time}
\label{sec:AccessTarget}
\noindent

The average of access time with respect to its  second index, the
{\it random target access time}  \cite{Lovasz:1993}, determines
the expected  number of steps a random walker needs to reach an
arbitrary node of the graph  chosen randomly from the stationary
distribution $\pi$ (if a random walk starts in $i\,\in\, V$). In
contrast to (\ref{sverage1}), the value of this average
  is independent of the starting node $i\,\in\, V$
being a {\it global}
spectral characteristic of a {\it graph},
\begin{equation}
\sum_{j\,\in\, V}\,
\pi_j\,H_{ij} \,
= \, \sum_{k=2}^N\,\frac 1{\,1-\mu_k\,}.
\label{target_spectral}
\end{equation}
 The latter equation expresses the so called
 {\it random target identity},
  \cite{Lovasz:1993}.
\begin{figure}[ht]
 \noindent
\begin{center}
\epsfig{file=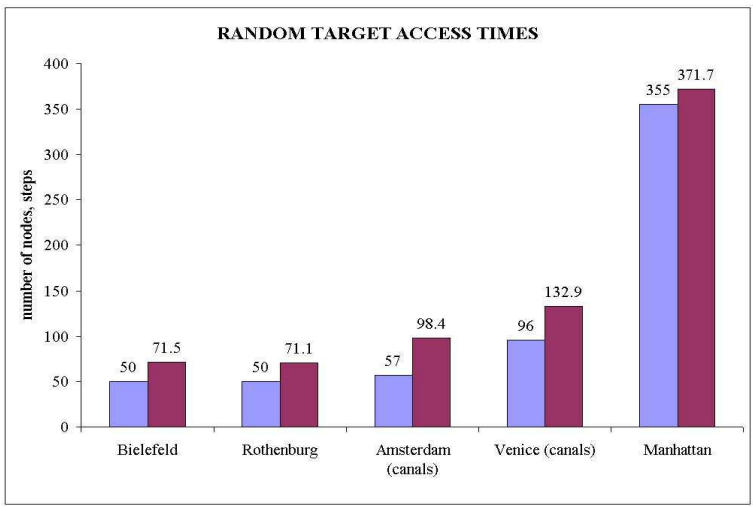,  angle= 0,width =10cm, height =6.7cm}
  \end{center}
\caption{\small The comparative diagram of the random target
access times and the sizes of dual graphs for five
 compact urban patterns. Heights of left pillars correspond to the number
 of nodes in the dual city graphs. Pillars on right
show the random target access times, the expected  number of steps a random
walker needs to hit a node randomly chosen from the stationary
distribution $\pi$.}
\label{Fig2_04a}
\end{figure}
Computations of random target
 access times (\ref{target_spectral})
for the dual city graphs  show that they
are closely tied to the network size
 (i.e., the numbers of
junctions between different streets and canals
in a city) (see Fig.~\ref{Fig2_04a}).
 The diagram shown on Fig.~\ref{Fig2_04a}  have been calculated for
 the dual graph representations of two German organic medieval cities
  founded shortly after the Crusades and developed within the medieval
   fortresses (Rothenburg ob der Tauber in Bavaria and the
   downtown of Bielefeld, nowadays the biggest city of
   Eastern Westphalia), the street grid in Manhattan (a borough
   of New York City with an almost regular grid-like city plan),
    and two city canal networks, in Amsterdam (binding to the delta
    of the Amstel river, forming a dense canal web exhibiting a high
     degree of radial symmetry) and in Venice (that stretches across
     122 small islands between which  the canals serve the function
     of roads), within the same frame.

\section{Conclusion}
\label{sec:Discussion}
\noindent

The properties of random walks
defined on the dual graph representation of a transport network is related to
the unique equilibrium configuration of not random commodity flows
along edges of its primary graph representation that unveils the
role Markov's chain processes and space syntax approach play in
the successful traffic modelling.
The amazing effectiveness of random walk models in describing
complex cooperative phenomena has been often discussed in
literature, \cite{Weiss:1994}. In space syntax theory, the concept of
random walks (implicitly reckoned by means of
 integration measures)
has been applied to the "probabilistic" analysis of spatial
configurations and proved its power by the surprisingly accurate
predictions of human behavior in cities \cite{Penn:2001}.

We have demonstrated that random walks establish the Euclidean
space structure
 on undirected graphs, in which the notions of distance and angle acquire
the clear statistical interpretations.
In particular, every open space (a street, a square, or a canal)
acquires a norm  based on
the quality of  path it provides for random walks.
This norm can be used as a measure of statistical segregation the node
is characterized in the given network.

In the present paper, we have canalized space syntax of Venetian's
canals by means of random walks defined on the dual graph
representation of the network and detect the segregated part that
is identified with the specific canal subnetwork of Venetian
Ghetto.

\section{Acknowledgment}
\label{Acknowledgment}
\noindent

The work has been supported by the Volkswagen Foundation (Germany)
in the framework of the project: "Network formation rules, random
set graphs and generalized epidemic processes" (Contract no Az.:
I/82 418). The authors acknowledge the multiple fruitful
discussions with the participants of the workshop {\it Madeira
Math Encounters XXXIII}, August 2007, CCM - CENTRO DE CI\^{E}NCIAS
MATEM\'{A}TICAS, Funchal, Madeira (Portugal).

\end{document}